\documentstyle[amssymb,amsmath,12pt]{article}

\def\slash#1{\setbox0=\hbox{$#1$}#1\hskip-\wd0\hbox to\wd0{\hss\sl/\/\hss}
}

\def\nd{\noindent}

\begin{document}

\begin{center}{\large\bf
A New Approach To The Hierarchy Problem\footnote{An extended version
of the talk 
presented during  
the activities of the European Meeting (CERN, 18-22 April 2001),
and at University of Bonn (29 May 2001),
based on hep-ph/0102307 \cite{Dvali:2001qr}.}
}
\end{center}
\centerline{
Rula
Tabbash\footnote{e-mail: rula@sissa.it}
}
\bigskip
\centerline{\it
SISSA--ISAS,
Via Beirut 4, I-34014 Trieste, Italy}

\vskip 1.2truecm

\begin{abstract}
We argue that 
identifying
the electroweak Higgs particle with the extra
components of the gauge field in $4+d$ dimensions provides a solution 
to the hierarchy problem. 
The absence of ultraviolate quadratic divergences is due to the fact
that 
the Higgs mass is protected by an exact gauge symmetry 
at energies beyond the cutoff. 
The idea is implemented within explicit models 
which also provide a link between fermion chirality
and electroweak symmetry breaking 
in four dimensions. 

\end{abstract}
\vskip 1truecm



\section{Introduction and Outline}

Despite the 
experimental success\footnote{In the recent past there has also 
been dissatisfaction
from the experimental point of view, since the Standard Model
is in shortage of explaining, for instance,  
neutrino oscillations and the 
recent data of $g-2$ of the Muon.} 
of the Standard Model as 
a theory describing the strong and electroweak interactions of 
elementary particles, this model 
is not theoretically satisfactory 
for various reasons.  
There are two main sources of 
theoretical
dissatisfaction; the first has to do with the model itself (when 
it comes to explaining, flavor, and charge quantization, for example), 
and
the second arises 
when the standard model is discussed 
within a more general context where the 
fourth fundamental force of nature, gravity, is present; 
leading subsequently to an instability of the 
weak scale at the quantum level. This latter obstacle, known as the 
{\it hierarchy problem} between the electroweak and gravity scales, 
was the main motivation for particle physicists to start searching 
for new physics beyond the standard model. These searches lead to 
the birth of many hoping-to-be-physical theories like technicolor, 
grand unified theories, supersymmetry, and recently models with 
large 
extra dimensions.

The Hierarchy problem is present at both classical and quantum levels.
At the tree level, there is a huge difference between the scales
associated with the electroweak and the gravitational interactions,  
$
{M_W}/{M_p}
\sim { 10^{-17}}$. Although this hierarchy is softened
when GUTs are considered, it persists to exist
$
{M_W}/{M_{GUT}}
\sim {10^{-13}}$. 

If quantum corrections were not to significantly alter the value of the 
Higgs (mass)$^2$ computed classically, one could consider the above mentioned 
hierarchy at the tree level as one of the many extreme ratios
existing in nature. However,
there are huge 
quadratic corrections to the Higgs
(mass)$^2$ at the quantum level
due to the fact that the Higgs particle is described by 
a fundamental scalar field. These corrections change the
classical value of the Higgs (mass)$^2$ by many orders of magnitude, and 
adjusting the value back to its classical one 
requires a fine-tuning of order  
${10^{-34}}$ in the case of a gravitational cutoff, and 
another of order ${10^{-26}}$ in GUTs. 

Supersymmetry is a very good example where the problem 
is nicely solved at the quantum level, in fact once the SUSY breaking scale 
is set classically, usually taken to be around few TeV, only logarithmic corrections 
will alter this value. The key principle for the absence of 
quadratic divergences in this theory is that the Higgs mass is protected by
the 
symmetry above the cutoff. Being in a multiplet with fermions, the Higgs
is deemed to be massless, due to  
chiral symmetry, as long as SUSY is unbroken.   

In the work 
\cite{Dvali:2001qr}, we employ the above key principle, though 
without supersymmetry, to explicit models where the Higgs mass is protected 
by a gauge symmetry in $4+d$ dimensions, the Higgs 
being identified with the 
components of a gauge field in higher dimensions.

The talk is organized as the following; in the beginning a short reminder 
of the original idea behind Kaluza--Klein theories, the idea on which 
our constructions are based,
is given. Then I will
describe 
in words the principle ideas we are proposing in our models, after
which 
I will move to discuss these models in more details. Firstly I 
will present 
a toy model for leptons in 6 dimensions, then 
a toy model for quarks and
leptons in 10 dimensions where Higgs mass is produced at the tree-level.
Then I will show how Higgs mass can be induced at the loop level, within
the same frame work. I will conclude with a summary and outlook. 


\section{In words}


\subsection{A historical remark}
Although having seven heavens is not a particularly new idea, 
the first  
to be published in a scientific journal, proposing our observed world to 
be an effective theory of a fundamental theory existing in more
than four dimensions,
was 
Nordstr\"{o}m's \cite{nord} back in 1914. Having no general relativity 
at that time, 
Gunnar
Nordstr\"{o}m wrote down Maxwell's equations in 5
dimensional space-time, and by wrapping the fifth dimension on a 
circle, he reduced the equations to Maxwell-Nordstr\"{o}m 
electromagnetic-gravitational theory in 4 dimensions. 

Five years later, in 1919, the mathematician 
Theodor
Kaluza
proposed \cite{kaluza} obtaining a four dimensional Einstein-Maxwell
theory starting from Einstein's gravity 
equations in five dimensions. Assuming the five dimensional manifold,
$W$,
to 
be a product of a 4 dimensional space-time $M_4$
and a circle $S^1$, $W=M_4\times S^1$, 
the fifteen components of the metric can be decomposed 
from 
a
four dimensional point of view 
into 
10
describing the 
gravity tensor,
four forming the components of a $U(1)$
gauge field, and one degree of freedom representing a scalar
field. By Fourier expanding those fields, retaining 
only the zero 
modes, 
and integrating along the circle $S^1$ one obtains
a theory in four dimensions which is invariant 
under both 
four-dimensional 
general coordinate transformations and abelian gauge transformations.  
and a massless vector boson which 
can be checked to 
have $U(1)$ gauge transformations.

In his original work, Kaluza assumed the zero mode of the 
scalar field to be constant, $
\phi^0=1
$. In any case, the value of $\phi^0$ had to be positive in order to insure
the proper relative sign of the Einstein and Maxwell terms so that the 
energy is positive. This in turn means that 
the fifth dimension must be space like. This can also be easily understood
in terms of causality; clearly a compact time-like dimension 
would lead to closed time-like curves.
The abelian gauge symmetry arising in four dimensions upon 
compactification originates
from the isometry of the circle. Those last two point, the requirement that
the extra dimensions must be space-like, and that the 
isometry of the compact space results in a gauge symmetry of the 
effective action are general arguments. 

Seven years later, Oskar Klein used Kaluza's idea in an 
attempt 
\cite{klein1}
to explain the underlying quantum mechanics of Schr\"{o}dinger 
equation by deriving it from a five-dimensional space
in which the Planck constant is introduced in connection with 
the periodicity along the closed fifth dimension. In this paper he 
also discusses the size of the compactified circle, getting 
closer to giving the extra dimensions a physical meaning than his 
predecessors.
In a separate 
work, still in 1926, Klein proposed\footnote{Credit
to this approach goes also to 
V. Fock and
H. Mandel, though others worked for 
and achieved
the same aim 
as well.
} a relativistic generalization of Schr\"{o}dinger's 
equation by starting from a massless wave equation in 
five dimensions and arriving at four-dimensional Klein-Gordon equation
for individual harmonics.

Afterwards,
many people adopted Klauza's idea, starting from Einstein early last 
century and continuing 
by several physicists today. During this period, the idea of having
new dimensions to propagate in inspired many to write the first 
complete
models for Lagrangians unifying Yang-Mills and gravity theories
\cite{kerner}, supergravity in 11 dimensions 
\cite{Cremmer:1978km}, and superstring theories which has to be 
considered in 10 dimensions for the theory to be 
anomaly free and hence consistent at the quantum level. 
In the original Kaluza--Klein framework, the particles are free 
to move in the compact space, as well as the to-be-observable
one, and hence the volume of the new dimensions should be small
in order not to undesirably interfere with the present 
observations and existing experimental data, since the new 
idea of $4+d$ dimensions has a strong potential to naturally
modify the expansion of the Universe and the cross sections of 
elementary particle interactions. 
Hence, the typical scale of compactification in string theory 
is taken to be of order of the four dimensional Planck 
length $M_P^{-1}$.

A recently activity of model building inspired by string theory, 
though not as mathematically rigorous, 
have been developed, and were launched by \cite{add}. The 
basic idea is still the one of Kaluza and Klein which considers a
tensor product (factorizable metric) of the four dimensional 
world and the compact space. 
Other models with 
infinite non-compact extra dimensions (with warp/non-factorizable 
metric) have also been 
phenomenologically considered \cite{Randall:1999ee}, while the original 
idea was proposed much earlier in \cite{Wetterich:1984uc}. 
The main aim of those models was to solve the 
hierarchy problem between the gravity and electroweak scales, by lowering
the fundamental scale of gravity in $4+d$ from $10^{19}$GeV down to
$1$TeV. In Kaluza--Klein type models, this requires assuming
a bigger compactification radii than the ones used in string theory, and
therefore new physics is expected to be observed at energies 
approaching the typical mass scale of compactification. In the 
warp compactification models, the extra space is non-compact. The 
exponentially damping warp factor considered in those 
models enters into the formulae linking the fundamental gravity scale
and the four dimensional one.    
Again, new dimensionful parameters
have to be
introduced and tuned in order to solve 
the hierarchy 
problem. 

So far no full
and consistent
model has been constructed, neither in tensor compactification nor 
in the warp one, however the phenomenological and cosmological 
implications have been studied extensively,
and attempts towards creating a theoretically appealing model 
persist.

\subsection{What we are doing}

The models we are interested in constructing are in the 
direction
of answering the following question: 
{\it ``How close can one get 
the to the Standard Model, in four dimensions, by Kaluza--Klein 
compactifying large extra dimensions of an Einstein Yang-Mills
theory coupled to fermions in $4+d$ dimensions?''}. 

Eventually, the effective action of a ``good'' theory should provide 
us in four dimensions with:
\begin{itemize}
\item{Chiral fermions.}
\item{The standard model gauge group, $SU(3)\times SU(2)
\times U(1)\rightarrow U(1)$, or a group containing it. } 
\item{Fermions and Higgs particles in the correct 
representations of standard model gauge group.}
\item{A good solution for the hierarchy problem.}
\item{Correct quarks and lepton masses.}
\item{Suppressed proton decay.}
\end{itemize}

We will give examples where the first four points mentioned above 
can be realized, with some hints to achieving the last two. 

In a an attempt to approaching the answer of the above question, we 
propose two principal ideas: 
\begin{itemize}
\item{Identifying the electroweak Higgs with the 
extra components of the Yang-Mills
field as a solution for the hierarchy problem.}
\item{Linking fermion chirality to the spontaneous electroweak 
symmetry breaking.}
\end{itemize}

\subsection{Approaching the hierarchy problem}
Our suggestion for the hierarchy 
problem is that the electroweak Higgs particle is identified with 
the extra components of the gauge field in $4+d$ dimensions. 
Suppose that we start from an Einstein Yang-Mills theory coupled to fermions
on a $4+d$ dimensional manifold, $W$, $W=M_4\times Y_d$. The field content
to start with  
consists of a graviton, $g_{MN}(x,y)$, 
a gauge field $A^a_M(x,y)$ in the algebra of a Lie group $G$, 
and a fermion $\psi(x,y)$. Where $M,N=0,1,2,...,d+3$, $a=1,...\mbox{dim}G$, 
$x$ and $y$ are the coordinates on $M_4$ and $Y_d$ respectively. We take 
$Y$ to be a compact manifold with a typical volume of order TeV$^d$, and 
$\times$ to indicate a tensor product.   
Arguing that the Higgs particle is  
$${H (x)\;\equiv \;A_\alpha(x)}\;\;\;\;\;\;\alpha\in Y$$
implies a solution for a hierarchy problem at the quantum 
levels, as was first pointed out by \cite{Hatanaka:1998yp}. 

Classically, the Planck scale is related to the fundamental 
$4+d$ dimensional gravity scale by the relation
$$
M_P= a^{\frac{d}{2}}\;M^{\frac{d}{2}+1}
$$
When $Y_d=Y_1\times Y_2\times...\times Y_n$, the $a^{d}$ 
should be replaced by the product of the volumes of each manifold.
As an example, we take 
$$
W=M_4\times S^2\times {\mathbb{C}}P^2
$$
as in the model we discussed in \cite{Dvali:2001qr} 
for quarks and leptons. In this example 
we have the gravity scale $M\sim 10^4$TeV, and hence new physics is 
expected to show off at around $1/a \sim 1$TeV.
Whether one considers the cutoff to be $M$, $1/a$, or the scale at 
which the gauge coupling in $10$ dimensions becomes strong 
(in our case this happens at around $1$TeV as well), the hierarchy 
between the weak scale and the cutoff, $\Lambda$, is much milder than 
the one in the ordinary gravity or grand unified theories: 
$$
\frac{m_H}{\Lambda} \sim 10^{-5} -- 10^{-1}
$$

Quantum mechanically, the absence of quadratic, or large, 
divergences can be understood via the argument of symmetry, as in the 
case of SUSY, however with no fundamental scalar to start with. 
The gauge symmetry in this case is spontaneously broken due 
to the presence of 
a topologically non-trivial background (as I will explain later on), 
however,
at energies larger than the compactification scale, 
it is recovered and the Higgs field is massless being
a component of the massless gauge field. 
In other words the Higgs mass$^2$, $m_H^2$, can not be larger 
than $1/a$  
$$m_H^2\;=\; \frac{1}{a^2}\;f(Ea)$$
\nd
where $E$ is the common energy scale, and $a$ is the typical radius 
of compactification. We conjecture that 
$\displaystyle
\lim_{E\to \infty}
f(Ea)=0$. In fact, it was shown in \cite{Hatanaka:1998yp} that 
the function $f(Ea)$ is exponentially damping at energies higher 
than $1/a$. Finding the explicit form of $f$ in our 
case is technically more complicated due to the presence of a  
monopole background, however we believe that the 
$\displaystyle
\lim_{E\to \infty}
f(Ea)$ will always be finite.  
 
\subsection{Linking chirality and SSB}
As it is well known, a topologically non-trivial background is 
needed in Kaluza--Klein field theories in order to get chiral fermions
in four dimensions as proposed by \cite{Randjbar-Daemi:1983hi} 
and discussed in \cite{Witten:ed.ux}.  
In the following examples we will show that the same monopole background
used to get chiral fermions is also responsible for the spontaneous
symmetry breaking (SSB) of the standard model gauge group.


\section{        
Monopole Background {$\boldsymbol {\Rightarrow}$} Chirality + SSB
}

Let us start from the following general action 
\begin{equation}
S= \int d^D x \sqrt{-G} (\frac{1}{\kappa^2}
{\cal R} + \lambda
-\frac{1}{2g^2}\mbox{Tr}F^2 +{\bar \psi } i \slash{\nabla} \psi  
+...)
\label{1}
\end{equation}
\nd
where $cal R$ is the Ricci scalar, $F$ is the field strength of the 
Yang--Mills potential, $\lambda$ is a cosmological constant, and 
$\psi$ is a fermion in some representation of the non-abelian 
gauge group $\cal G$. The dots imply all 
possible higher order non-renormalizable 
operators. 

In order to illustrate the idea, let us begin by a simple example 
of leptons in six dimensions. 

\subsection{$\boldsymbol{SU(3)}$ in $\boldsymbol{D=6}$
}
The bosonic Einstein and Yang--Mills equations of motion 
derived from the action
(\ref{1}) admit the following solutions; for the metric
$$
ds^2=\eta_{\mu\nu}dx^\mu dx^\nu + a^2(d\theta^2+\mbox{sin}\theta 
d\varphi^2)
$$
(in other words $W=M_4\times S^2$ is a solution), and
\begin{equation}
\left<A_\varphi\right> = 
\frac{1}{2} H (\mbox{cos}\theta-1)d\varphi
\label{2}
\end{equation}
for the background gauge field.\footnote{We show the calculations
performed
on the upper hemisphere only for simplicity. However,  
the same can be done, with consistent patching, 
on the 
lower hemisphere, as well, by starting from the solution 
$
\left<A_\varphi\right> = 
\frac{1}{2} H (\mbox{cos}\theta+1)d\varphi
$.} Where 
$H=\mbox{diag}(n_1,n_2, -n_1-n_2)$, is in the Cartan subalgebra of 
$SU(3)$, and $\;n_1,n_2$
have to be integers 
in order for the fermions to patch 
properly on the upper and lower hemispheres.

It is clear that the monopole background (\ref{2})
breaks the $SU(3)$ symmetry to $U(1)\times
\tilde
U(1)$ for a generic $n_1$ and $n_2$. 
The metric is also invariant under the rotation of
the
$S^2$ covered by $\theta$ and $\varphi$. Thus for arbitrary
values
of $n_1$ and $n_2$ the effective d=4 theory will be
invariant
under $SU(2)\times U(1)\times \tilde U(1)$. In case that
$n_1=
n_2 = n$, or its equivalent\footnote{
For example, the case $(n_1,n_2)=(1,1)$ is equivalent to 
$
(n_1,n_2)=(-1,2)
$.
}, the maximal $SU(2)\times  U(1)$ 
will remain unbroken thereby enlarging the 4 dimensional symmetry to 
$SU(2)\times  SU(2)\times \tilde U(1)$. We are going to adopt this
second case, and shall show that the two 
$SU(2)$'s (call them as $SU(2)_L$ and $SU(2)_R$, indicating the 
isometry group and the subgroup of $SU(3)$ respectively) 
act chirally.

Let us start by computing the fermion spectrum. 

\subsubsection{Spectrum of chiral fermions}
The Dirac equation on a generic manifold is 
$$
\slash{D}_W\psi(x,y)=0
$$
Suppose $W=M_4\times S^2$, the above equation can be 
written, with the
appropriate choice of $\Gamma$ matrices, as  
$$
\slash{D}_4\psi(x,y)+ \slash{D}_{S^2}\psi(x,y) =0 
$$
Note that in the absence of a background gauge field
\begin{itemize}
\item{$\slash{D}_{S^2}\psi=0 $ has no
{\it regular} solutions.}
\item{ $\mbox{ Index}\slash{D}_{S^2}=0$.}
\end{itemize}
The first point can be understood easily, with and without 
going into explicit 
computations, since, 
according to Lichnerowicz's 
theorem, 
all positively curved 
smooth compact manifolds, including $S^2$,
do not admit harmonic spinors. 
In order for the resulting fermion in 4 dimensions to be 
chiral, it should be massless to start with. 
Hence the first requirement
for the Dirac operator on the compact internal manifold
is to have at least one zero mode. Also, the index of the 
Dirac operator is often, not always, zero.

The only way to get a solution for $\slash{D}_{S^2}\psi=0 $ is to 
couple it to a topologically non-trivial background \cite{
Randjbar-Daemi:1983hi,Witten:ed.ux}. This also changes  
$\mbox{ Index}\slash{D}=0$ and hence allows for achieving a chiral theory, 
as the standard model, in four dimensions. 

Consider the Dirac operator on $S^2$ coupled to the monopole background
(\ref{2}):
$$
\slash{D}_{S^2}=
\Gamma^mE_m^\alpha(\partial_\alpha
-\frac{1}{2}\omega_{\alpha [k,l]} \Sigma^{kl}-i
\left<A_\alpha(y)\right>)
$$
The background solution is of the form:
\begin{equation}
\left<A_\alpha(y)\right> = \frac{1}{2}H \omega_\alpha
\label{eq}
\end{equation}
\nd
where $\omega_\alpha$ ($\alpha =1,2$) is;
$\omega_\theta =0$, and $\omega_\varphi = -(\mbox{cos}
\theta -1)$.

Now let $\psi(x,y)$ be a Weyl spinor in 6 dimensions, in the 
${\underline{3}}$ of $SU(3)$. Using the chirality matrix, $\gamma_5$,
in four dimensions, $\psi$ can be written as: 
$$
{\psi}= \frac{1+\gamma_5}{2}\psi+\frac{1+\gamma_5}{2}\psi
={\psi_R+\psi_L}
$$
The Dirac equation in a monopole background hence simplifies to
\begin{eqnarray}
\partial_\theta {\psi_R}+\frac{i}{\mbox{sin}\theta}
{\psi_R}
+{\lambda_R\psi_R} \;\mbox{ctg}\theta
&=&0
\nonumber\\
\boldsymbol{-}\partial_\theta {\psi_L}+\frac{i}{\mbox{sin}\theta}
{\psi_L}
+{\lambda_L\psi_L} \;\mbox{ctg}\theta
&=&0
\nonumber
\end{eqnarray}
\nd
where the ``isohelicities''\footnote{See 
\cite{Randjbar-Daemi:1983bw}.} 
of the various components of 
the fermions can be tabulated in
$$ 
{\lambda}({\psi_R}) =
\left(\begin{array}{ccc}
\frac{1}{2} (1+n_1)&  \\[5pt]
\frac{1}{2}(1+n_2)  &\\[5pt]  
\frac{1}{2} (1-n_1-n_2) & 
\end{array}\right)
$$
and
$$
{\lambda}({\psi_L}) =
\left(\begin{array}{ccc}
\frac{1}{2} (-1+n_1)&  \\[5pt]
\frac{1}{2} (-1+n_2)  &\\[5pt]  
-\frac{1}{2} (1+n_1+n_2) & 
\end{array}\right)
$$

It can be 
checked that regular solutions exist only for 
$\lambda_l \geq
0$ and $\lambda_R \leq
0$.

For instance, adopting the choice
$(n_1, n_2) =
(-1,2)$, the spectrum consists of two right-handed singlets of the 
Kaluza--Klein $SU(2)_L$ and one left-handed doublet of the
same $SU(2)_L$. Including the transformation under $SU(2)_R$ and $\tilde
U(1)$, we can summarize the chiral fermion spectrum  
as 
$({2}_L,{1}_L)_{1}+
({1}_R,{2}_R)_{1/2}$ where the subscripts indicate the 
$\tilde U(1)$ charges.      

\subsubsection{Higgs spectrum}
Having explained above how coupling to a monopole background can 
result in chiral fermions upon compactification, the aim 
now is to show how coupling to the same background derives spontaneous
symmetry breaking. 

To start with, there are two sources for the d=4 boson fields, those
originating
from the fluctuations of the metrical backgrounds and those
stemming from the d=6 gauge field fluctuations. In general
the
effective four dimensional fields are linear combination of
these
two types of fluctuations. The modes we are interested in,
however, do not mix with gravity. Therefore in this paper we
shall not include the gravitational fluctuations. The
bosonic part of the
effective 4-dimensional theory is then obtained by expanding
the
6-dimensional Yang-Mills action on the background of our
classical solution.
 
We shall write 
$$
A_C(x,y)= \left< A_C(y)\right> + V_C(x,y) $$
\nd
where $\left< A_\mu(y)\right>=0$, $\mu = 0,1,2,3$, 
and 
$\left< A_\alpha(y)\right>\neq 0$
is the background background monopole 
configuration given in (\ref{2}). For the gauge field
fluctuations $V_\mu$ we shall retain only the ones
in the algebra of the unbroken gauge group, while for the fields
$V_\alpha$ tangent to $S^2$, we shall retain
the components in the complementary subspace in the Lie algebra. It is
only these ones which become tachyonic in the monopole background and hence
lead to the spontaneous symmetry breaking.

Substitute the above in the Yang-Mills field strength,
$$
F_{AB}= \partial_A A_B - \partial_B A_A -i[A_A, A_B]= 
D_A V_B - D_B V_A -i[V_A, V_B]\ .
$$
where $D_A V_B$is defined by
$$
D_A V_B = \nabla_A V_B -i[A_A, V_B]\ .
$$
\nd
where $\nabla_A V_B$ represents the 
ordinary Riemannian derivative.

As has been explained in detail in 
\cite{Randjbar-Daemi:1983hi} and \cite{Randjbar-Daemi:1983bw}
in performing the harmonic expansion on $S^2$  it is convenient to
use the $U(1)$ basis on the tangent space to $S^2$. The mode structure of each 
field is then completely fixed by its "isohelicity" $\lambda$.
The isohelicities can be tabulated as  
\cite{Randjbar-Daemi:1983bw} 
$$
\lambda(V_a) =
\left(\begin{array}{ccc}
0                      & {1\over 2} (n_1-n_2)   & {1\over 2} (2n_1+n_2) \\[5pt]
-{1\over 2} (n_1-n_2)  & 0                      & {1\over 2} (n_1+2n_2) \\[5pt]
-{1\over 2} (2n_1+n_2) & -{1\over 2} (n_1+2n_2) & 0
\end{array}\right)
$$
$$
\lambda(V_\pm) =
\left(\begin{array}{ccc}
\pm 1                       & {1\over 2} (n_1-n_2) \pm 1  & {1\over 2} (2n_1+n_2)\pm 1 \\[5pt]
-{1\over 2} (n_1-n_2)\pm 1  & \pm 1                       & {1\over 2} (n_1+2n_2)\pm 1 \\[5pt]
-{1\over 2} (2n_1+n_2)\pm 1 & -{1\over 2} (n_1+2n_2)\pm 1 & \pm 1
\end{array}\right)\ \ \ .
$$

The harmonic expansion on $S^2$ then proceeds 
as in \cite{Randjbar-Daemi:1983hi} and \cite{Randjbar-Daemi:1983bw}, viz,
$$
\begin{array}{rcl}
V_a(x,\theta,\varphi) & = & \sum\limits_{1\geq \vert\lambda\vert} \sum\limits_m ({2l+1\over 4\pi})^{1/2}
D^l_{\lambda,m}(\varphi,\theta)V^l_{am}(x)\ ,\\[5pt]
V_\pm (x,\theta,\varphi) & = & 2^{-1/2}(V_4\mp i V_5)\\[5pt]
 & = & \sum\limits_{1\geq\vert\lambda\pm \vert}  \sum\limits_m
({2l+1\over 4\pi})^{1/2}D^l_{\lambda\pm,m}(\varphi,\theta)V^l_{\pm m}(x)\ ,
\end{array}
$$
\nd
where $D^l_{\lambda m}$ are the $SU(2)$ -representation matrices
with Euler angles $\phi, \theta,$ and $ \pm\phi$.

The linearized Yang-Mills equations, in 
the covariant gauge $D_AV_A=0$ satisfied by the component
fields $V^l_{bm}(x)$, $V^l_{\pm m}(x)$ are
$$
\{\partial^2 -{ a^{-2}
[l(l+1)-\lambda^2]}\} {V_b} = 0, \;\;\; l\geq \;|\lambda|
$$
$$
\{\partial^2 - {a^{-2}[l(l+1)-(\lambda_\pm 
\mp 1)^2]}\} {V_\pm } = 0, \;\;\; l\geq
\;|\lambda_\pm| 
$$
where $a$ is the radius of $S^2$.

The components $V_\pm$ clearly exhibit tachyons. 
For example, in $V_+$ the component
$l=\vert\lambda_+\vert$ for $\lambda_+\leq 0$ carries negative 
(mass)$^2$ \cite{Randjbar-Daemi:1983bw}
$$
-(\vert\lambda_\pm\vert + 1)/a^2\ .$$

As we saw while deriving the fermionic spectrum, it is necessary 
to chose $(n_1,n_2)=(-1,2)$, or its equivalent,
in order to obtain left handed chiral 
fermions in the doublet representation of 
$SU(2)_L$.
It is an interesting accident that for 
the same choice of the magnetic charges the leading term in the 
expansion of $V_+ ^6 + iV_+ ^7$  is tachyonic and has 
$l=1/2$.  We thus obtain a d=4 Higgs doublet of $SU(2)_L$. 

Retaining only the massless gauge fields and this tachyonic field 
and integrating over the coordinates of
$S^2$ we obtain the following 4 dimensional effective action:
\begin{eqnarray}
{\cal L}_B&=& -\frac{1}{2g^2}
\int_{0}^{2\pi}d\varphi\int_{0}^{\pi} d\theta \;
\mbox{sin}\theta \;\mbox{Tr} F_{MN}F^{MN}
\nonumber\\
&=& -\frac{1}{4g_1^2} {F^8_{\mu\nu}}^2 -
\frac{1}{4g_2^2} {F^r_{\mu\nu}}^2
-\frac{1}{4e^2} {W^r_{\mu\nu}}^2
\nonumber\\
&-&\mbox{Tr}\left\{
|(
\partial_\mu-\frac{3}{2}iV_\mu^8
-iV_\mu^r \frac{\sigma^r}{2}-i W_\mu^r 
\frac{\tau^r}{2})\phi
|^2
-\frac{3}{2a^2}
\phi^\dag\phi+2g_1^2 (\phi^\dag\phi)^2\right\}
\nonumber
\end{eqnarray}
\begin{eqnarray}
{\cal L}_F&=&
\int_{0}^{2\pi}d\varphi\int_{0}^{\pi}d\theta\;
\mbox{sin}\theta\; {\bar \psi}i \slash{\nabla}\psi
\nonumber\\
&=&{\bar \lambda}_L\; i \gamma^\mu\left(
\partial_\mu-ig_1V_\mu^8-ie W_\mu^i\frac{\tau^i}{2}
\right)\lambda_L
\nonumber\\
&+&
{\bar \lambda}_R\; i \gamma^\mu\left(
\partial_\mu-i\frac{g_1}{2}V_\mu^8-ig_2 V_\mu^i
\frac{\sigma^i}{2}\right)\lambda_R\nonumber\\
&
-&2g_1
\left\{{\bar \lambda}_L\phi (i\sigma_2)\lambda_R
-{\bar \lambda}_R (i\sigma_2)\phi^\dag\lambda_L
\right\}
\nonumber\\
\end{eqnarray}
\nd
where $W^i_\mu$ indicate the Kaluza--Klein $SU(2)_L$
gauge potential and $W_{\mu\nu}^i $ is its corresponding 
field 
strength. $e$ denotes the Kaluza--Klein gauge coupling  constant. The
Higgs doublet $\phi$, with some appropriate constant rescalings,
is  the $x$-dependent coefficient of the $l=1/2$ term 
in the expansion of  $V_{+} ^6 + iV_+ ^7$.  
The $SU(2)_R\times\tilde U(1)$
couplings $g_2$ and $g_1$ are 
related to the six dimensional gauge coupling and the radius
of $S^2$ via Einstein equations
$$
g_2= {1\over 2\sqrt \pi} {g\over a} = \sqrt 3\   g_1\; .
$$

With above choice of magnetic charges, the components
$V_+^4 +iV_+^5$ and $V_+^6 +iV_+^7$ form an $SU(2)_R$ doublet which 
is charged under the $\tilde U(1)$. Again its $l=1/2$ harmonics on
$S^2$ is tachyonic. 
As can be seen, the Higgs field transforms as
$(2,2)_{3\over 2}$. The chiral fermions are
$\lambda_L=(2_L,1)_1$,  and $\lambda_R=(1,2_R)_{1/2}$.

\subsection{$\boldsymbol{U(6)}$ in $\boldsymbol{D=10}$
}\label{iso}
Solution for the bosonic equations of motions are: 
$$
ds^2=a_1^2\left(d\theta^2+\mbox{sin}^2\theta d
\varphi^2 \right)
+\frac{4a_2^2}{1+\zeta^{\dag}\zeta}
d{\bar \zeta}^a \left(\delta^{ab}-
\frac{\zeta^a{\bar\zeta}^b}{1+\zeta^{\dag} \zeta}
\right)d\zeta^b
$$
\nd
for the metric, or in other words
$$
{W=M_4\times S^2  \times {\mathbb{C}}P^2},$$
\nd
and 
\begin{equation}
{\left<\;A(y)\;\right> = \frac{n}{2}(\mbox{ cos } \theta-1)d\varphi 
+ qi\omega
\label{back}
}
\end{equation}
\nd
where $(\mbox{cos}\theta-1)d\varphi$ and $\omega$ are the connections
on $S^2$ and $
{\mathbb{C}}P^2
$ respectively. 
$$
\omega(\zeta, {\bar \zeta})= \frac{1}{2\left(1+\zeta^{\dag}
\zeta\right)}\left(\zeta^\dag d\zeta - d\zeta^\dag \zeta
\right)
$$
\nd
$d\omega$ is the self dual K\"{a}hler form on
${\mathbb{C}} P^2$. It is thus an instanton type solution of
the Yang-Mills equation in ${\mathbb{C}} P^2$ (remember that the 
first term is nothing but the monopole solution (\ref{2}) on 
$S^2$). The coupling of the ${\mathbb{C}} P^2$ instanton   
to the fermions is crucial in defining spinors globally 
on ${\mathbb{C}} P^2$.

We chose the matrices $n$ and $q$ to be
${n=\mbox{diag}(-2,+1,+1,-2,+1,+1)}
$, and 
${q=\mbox{diag}(+\frac{5}{2},+\frac{5}{2},
+\frac{5}{2},+\frac{3}{2},+\frac{3}{2},+\frac{3}{2})
}$ (again $n$ has to be an integer while $q$ a half of 
an odd integer for topological reasons).

The procedure to adopt is very similar to the previous model 
for leptons explained before, however the computations here are
more subtle. In order to cut the long story short, I will give here
only the results. 

The vacuum in 4 dimensions is invariant under the isometry 
groups of both $S^2$ and ${\mathbb{C}} P^2$ which are 
$SU(2)$ and $SU(3)$ respectively. 
We start from a Weyl spinor in the \underline{6} of $U(6)$ in 10 dimensions.
The fermion spectrum we obtain in 4 dimensions can be 
summarized as 
$$
({2_L},{\underline{3}})+
({1_R},{\underline{3}
})+
({1_R},{\underline{3}})+
({2_L},{1})+
({1_R},{1})+({1_R},{1})
$$
$$
\;
\downarrow\;\;\;\;\;\;\;\;
\;\;
\;\;\;
\downarrow\;\;\;\;\;\;\;\;
\;\;\;\downarrow\;\;\;\;\;\;\;\;\;
\;
\;\;\;\;
\downarrow\;\;\;\;\;\;\;\;\;\;\;\;
\downarrow\;\;\;\;\;\;\;\;\;\;\;\;
\downarrow\;\;\;
$$
$$
\left(
\begin{array}{c}
u\\ d
\end{array}\right)_{{L}}\;\;\; 
\;\;
{u_R}\;\;\;\;\;\; \;\;\;\;
{d_R} 
\;\;\;\;\;\;
\left(\begin{array}{c}\nu_e\\ e
\end{array}\right)_{{L}}\;\;\;\;{e_R}\;\;\;\;\;\;\;\;\;\;
\psi_R
\;\;\;
$$
where the arrow are pointing towards the corresponding degrees of 
freedoms in the standard model. The additional right handed singlet
can be removed by an appropriate choice of $q$. The Higgs spectrum 
consists of two doublets of $SU(2)$, 
$
{\phi_1}= ({2},{1})$, and 
${\phi_2}= ({2},{1})$, which are singlets of $SU(3)$.

The vacuum is also invariant under the left unbroken subgroup 
of $U(6)$ which is
$SU(2)\times SU(2)\times U(1)^3$. We assume in \cite{
Dvali:2001qr} that this group is weakly coupled while the standard model 
group originates from the isometry group. 
Nevertheless, we include the 
transformation properties of the fermions under this group:
\begin{eqnarray}
(2_L,\underline{3})&\sim&(1,1)_{(-2,0,1)}\nonumber\\
(1_R,\underline{3})+(1_R,\underline{3})
&\sim& (2,1)_{(1,0,1)}\nonumber\end{eqnarray}
$$
(2_L,1)\sim  (1,1)_{(0,-2,-1)}\;\;\;\mbox{and}\;\;\;
(1_R,1)\sim  (1,2)_{(0,1,-1)}
$$

The Higgs doublets, on the other hand, have the transformations
$\phi_1= (2,1)_{(-3,0,0)}$, and $\phi_2 = (1,2)_{(0,-3,0)}$.

This model is free of global anomalies, and can be made 
free of local anomalies as well \cite{Dvali:2001qr}. 

These two models can not be considered 
realistic the way they stand. One of the reasons is that the mass 
of fermions in four dimensions are of the order of the 
compactification scale. This would not have been a problem 
by itself since those fermions could in principle be identified 
with the third generation. The problem is that, even in this case, there
is no justification for ignoring the higher Kaluza--Klein 
modes which also have masses of the same order. 
Another reason is that 
the resulting group in four dimensions
is bigger than the standard model group, and the assignments of 
hypercharges are not correct.  

\section{Loop induced Higgs mass }

As a way to overcome the issue of incompatible scales, the first
thing which comes to mind is to produce the Higgs mass by quantum 
corrections. The idea is to construct a model where Higgs mass is 
zero at the tree level.
The sign of the one loop induced effective
mass will depend on the imbalance between the contribution of
fermions and bosons. By a judicious choice of the fermionic
degrees of freedom this sign can be made tachyonic.
We proceed by showing that a zero tree level Higgs mass is possible in a 
monopole background by considering 
${U(N)}$ gauge theory in ${D=10}$ compactified on three spheres
$W=M_4\times S^2\times S^2\times S^2$ (with radii 
$a$, $a'$, and $a''$ respectively),
with three monopoles
on each sphere. The ansatz for the background gauge filed is 
$$
{A} = {{ {n}}\over 2}(\mbox{cos} \theta -1)d\phi + 
{{{n'}}\over 2}(\mbox{cos} \theta'
-1)d\phi' +{{{n''}}\over 2}(\mbox{cos} \theta'' -1)d\phi''
$$
\nd
where $n,n',n''\subset U(N)$, and they are $N\times N$ diagonal real
matrices. 
The scalars of interest for us are those components of the
fluctuations of the vector potential which are tangent to
$S^2\times S^{'2}\times S^{"2}$ and are in the directions of
perpendicular to the Cartan subalgebra of $U(N)$. Consider the
field $V_{-i}^{j}$ tangent to $S^2$. The masses of these
fields are related to the eigenvalues
of the Laplacians 
on the three spheres which are, in 
turn, determined from the
isohelicities of $V_{-i}^{j}$
$$
\lambda = -1 +{1\over 2}(n_i-n_j),\;
\lambda^{'} = {1\over 2}(n_i^{'}-n_j^{'}),\;
\lambda^{''} ={1\over 2}(n_i^{''}-n_j^{''}).$$
For illustration, we 
consider an 
$n$ matrix which has only the 
elements $n_1$ and $n_2$ different from zero and
such that $n_1 -n_2 \geq2$. Then $\lambda (V_{-1}^{2})\geq 0$ and
according to our general rule the leading mode in this field can
be tachyonic.
The question we would like to answer is if by an
appropriate choice of magnetic charges we can make the mass of
this field to vanish. The answer, as shown in \cite{Dvali:2001qr}, 
is yes. 
The masses of the infinite tower of modes of
$V_{-1}^2$ are given by
$$a^{2}M^2= l(l+1) - \lambda^2 + {a^2\over a^{'2}}
(l'(l'+1) -\lambda^{'2})
+ {a^2\over a^{''2}}
(l''(l''+1) -\lambda^{''2}) + 1-(n_1-n_2)$$
and by employing the bosonic background equations, and making 
a specific choice for the magnetic charges, 
the leading mode is indeed massless \cite{Dvali:2001qr}. 
For the same choice there will
of course be a similar massless mode in the fluctuations $V_{-
1}^{'2}$ tangent to $S^{'2}$. 
One can make all
other modes to have positive masses by appropriate choices of the
remaining magnetic charges.

\section{Summary and Outlook}
We argued that the electroweak Higgs scalar field originates from the 
extra components of a Yang-Mills potential in $4+d$ dimensions.
No $\Lambda^2$ divergences are expected to be present at high energies, simply 
because the Higgs will be a component of a massless gauge 
boson at energies beyond $1/a$. The computation of the Higgs mass
in low energies is basically the one of vacuum polarization in 
$4+d$ dimensions, after integrating over the compactified space 
and restricting the on-shell momenta to low energies.  
 
This scalar induces spontaneous symmetry breaking  
in 4 dimensions due to the 
presence of 
a magnetic monopole like background.
The same background produces chiral fermions in 
4 dimensions, hence linking both the 
electroweak symmetry breaking and fermion 
chirality in the standard model to have one 
source: the monopole background, and/or the instanton, 
in the internal space.

Chiral fermions obtained are in the 
correct 
representations of
${SU(3)\times SU(2)}$. Two models were worked out, where 
the standard model gauge group is assumed to be stemming
from the isometry groups of the compactified space
which are $SU(2)$ and $SU(3)$ for 
$S^2$ and 
${\mathbb{C}}P^2$ respectively (see section \ref{iso}). 
the first was a toy model for leptons, using  
an ${SU(3)}$ gauge theory in 6 dimensions compactified 
on an ${S^2}$, in the 
presence of a  
monopole. 
The second was a 
toy model for quarks and leptons, with a 
${U(6)}$ gauge theory  
in 10 dimensions on ${S^2\times {\mathbb{C}}P^2}$, 
in the background of a  
monopole and an instanton. In the latter model, the 
tachyonic mode, to be identified with the Higgs, 
is a singlet of $SU(3)$, and hence only $SU(2)$
gets spontaneously broken. 

These models can not be considered realistic as they stand 
for several reasons. 
Mainly the
hypercharge assignments are incorrect, and the resulting 
group in four dimensions is larger than the standard model one.  
Besides, the 
masses of quarks and leptons are too big
$\sim {\cal O}(1/a)$ (no justification for ignoring  
KK modes which have masses of the same order).

The problem of fermion mass scales can be overcome
by producing the Higgs mass at the quantum level.
An existence example was given; a 
$U(N)$ Yang-Mills in 10 dimensions compactified on three 
spheres in the background of three monopoles, where the 
Higgs mass is zero at the tree level and is expected 
to develop a tachyonic mass at one loop due to the 
imbalance between 
the bosonic and fermionic degrees of freedom. A work in progress
may reveal some further 
phenomenological 
aspects of the idea. 
\\


\nd
{\bf Acknowledgment}\\
I am grateful to Seif Randjbar-Daemi  
for many instructive and enlightening discussions. 
This work is partially supported
by the European TMR project ``Supersymmetry And The Early Universe'' 
under the contract HPRN-CT-2000-0152.


\begin{thebibliography}{99}

\bibitem{Dvali:2001qr}
G.~Dvali, S.~Randjbar-Daemi and R.~Tabbash,
%
``The origin of spontaneous 
symmetry breaking in theories with large  extra dimensions,''
hep-ph/0102307.

\bibitem{nord}
G.~Nordstr\"{o}m, Phys. Zeitsch. {\bf 15} (1914) 504.

\bibitem{kaluza}
Th.~ Kaluza, Sitzungsber. Preuss. Akad. Wiss. Phys. Math. Klasse 966 
(1921).

\bibitem{klein1}
O.~ Klein, Nature {\bf 188} (1926) 516.

\bibitem{klein2}
O.~ Klein, Z. F. Physik {\bf 37} (1926) 895.

\bibitem{kk}
The main source of the historical introduction is:\\
``Modern Kaluza--Klein Theories'', (Edited by T.~ Appelquist, 
A.~ Chodos, and P.~ Freund), Frontiers In Physics v. 65, Addison-Wesley 
Pub. Com. (1987).


\bibitem{kerner}
R.~ Kerner, Ann. Inst. H. Poincar\'{e} {\bf 9} (1968) 143.


\bibitem{Cremmer:1978km}
E.~Cremmer, B.~Julia and J.~Scherk,
Phys.\ Lett.\ B {\bf 76} (1978) 409.

\bibitem{add}
N.\ Arkani-Hamed, S.\ Dimopoulos, G.\ Dvali,
Phys. Lett. {\bf B429}\ (1998) 263;
Phys. Rev. {\bf D59}\ (1999) 086004;
I. Antoniadis, N. Arkani-Hamed,
S. Dimopoulos, G. Dvali, Phys. Lett. {\bf B436}\
(1998) 257.
Earlier suggestions to assign a
physical meaning to extra dimensions were made by 
K.~Akama,
Lect.\ Notes Phys.\  {\bf 176} (1982) 267;
V.~A.~Rubakov and M.~E.~Shaposhnikov,
Phys.\ Lett.\ B {\bf 125} (1983) 136.


\bibitem{Wetterich:1984uc}
C.~Wetterich,
Nucl.\ Phys.\ B {\bf 242} (1984) 473;
S.~Randjbar-Daemi and C.~Wetterich,
Phys.\ Lett.\ B {\bf 166} (1986) 65.

\bibitem{Randall:1999ee}
L.~Randall and R.~Sundrum,
Phys.\ Rev.\ Lett.\  {\bf 83} (1999) 3370;
L.~Randall and R.~Sundrum,
Phys.\ Rev.\ Lett.\  {\bf 83} (1999) 4690.

\bibitem{Hatanaka:1998yp}
H.~Hatanaka, T.~Inami and C.~S.~Lim,
Mod.\ Phys.\ Lett.\ A {\bf 13} (1998) 2601;
H.~Hatanaka,
Prog.\ Theor.\ Phys.\  {\bf 102} (1999) 407;
Y.~Hosotani,
Annals Phys.\  {\bf 190} (1989) 233.

\bibitem{Randjbar-Daemi:1983hi}
S.~Randjbar-Daemi, A.~Salam and J.~Strathdee,
Nucl.\ Phys.\ B {\bf 214} (1983) 491.

\bibitem{Witten:ed.ux}
E.~Witten,
``Fermion Quantum Numbers In Kaluza--Klein Theory'',
The Proc. of Second Shelter Island Meeting (1983) 227.

\bibitem{Randjbar-Daemi:1983bw}
S.~Randjbar-Daemi, A.~Salam and J.~Strathdee,
Phys.\ Lett.\ B {\bf 124} (1983) 345.

\end{thebibliography}
\end{document}